\documentclass[sigconf, nonacm]{acmart}
\usepackage{amsmath}
\usepackage{multicol}
\usepackage{dirtytalk}
\usepackage[super]{nth}
\usepackage[linesnumbered,ruled,vlined]{algorithm2e}
\DeclareMathOperator*{\argmax}{argmax}
\usepackage{etoolbox}
\makeatletter
\preto{\@verbatim}{\topsep=4pt \partopsep=4pt }
\makeatother

\begin{document}
\title{Deep Learning Framework From Scratch Using Numpy}

\author{Andrei Nicolae}
\affiliation{%
  \institution{Microsoft Corp.}
  \city{Bellevue}
  \state{Washington}
  \postcode{98004}
}
\email{andrei.nicolae@microsoft.com}

\begin{abstract}
This work is a rigorous development of a complete and general-purpose deep learning framework from the ground up. The fundamental components of deep learning - automatic differentiation and gradient methods of optimizing multivariable scalar functions - are developed from elementary calculus and implemented in a sensible object-oriented approach using only Python and the Numpy library. Demonstrations of solved problems using the framework, named ArrayFlow, include a computer vision classification task, solving for the shape of a catenary, and a 2nd order differential equation.
\end{abstract}

\maketitle

\begingroup
\renewcommand\thefootnote{}\footnote{\noindent
Alternate email address: nandrei@uw.edu}
\endgroup

\begingroup\small\noindent\raggedright\textbf{Source Code and Demos:} \\
\url{https://github.com/a-nico/ArrayFlow}
\endgroup

\section{Introduction}
The most effective method of developing a thorough understanding of an algorithm is by implementing it from scratch. In the process, obscure but significant facets are expounded which would otherwise be overlooked in a superficial treatment of the subject. Accordingly, ArrayFlow was developed from first principles with full originality and without any surveying of source code, papers, books, or the like - doing so is akin to examining the solution of a learning exercise before attempting to solve it, which would render the effort futile. 

Although ArrayFlow is capable of training network models of arbitrary architecture, size, and complexity, the purpose is not to replace or add to existing frameworks such as PyTorch and TensorFlow, but rather, as an enterprise to develop the details and nuances of deep learning. The framework name \say{ArrayFlow} is a jab at TensorFlow to remark that deep learning does \emph{not} deal with tensors but merely arrays of real numbers. This paper is best read in the company of the Demos notebook and source code made available on GitHub.

\section{Fundamentals}

\subsection{The Gradient}

Let $\phi: \mathbb R^n \to \mathbb R$ be a differentiable scalar field. Let 
\begin{align}
    \{ \mathbf u \in \mathbb R^n  \ | \ \left| \mathbf u \right| = 1 \}
\end{align}
denote the set of all unit vectors in the domain space of $\phi$.
Given a point $\mathbf x_0 \in \mathbb R^n$ suppose we wish to find another point $\mathbf x_1$ such that
\begin{align} \label{optimgoal}
    \phi(\mathbf x_1) > \phi(\mathbf x_0)    
\end{align}
The best linear approximation to $\phi$ at $\mathbf x_0$ is given by the linear terms of the Taylor series \cite{taylor}
\begin{align}
    \phi(\mathbf x + t \mathbf u) \approx \phi (\mathbf x) + t \mathbf u \cdot \nabla \phi
\end{align}
provided that $\nabla \phi$ exists and is continuous in the $t$ neighborhood of $\mathbf x$. Given this information, the best guess for $\mathbf x_1$ is $\mathbf x_0 + t \mathbf u$
\begin{align}
    \argmax_{\mathbf u} \lim_{t \to 0} \phi(x + t \mathbf u) = 
    \frac{\nabla \phi(\mathbf x)}{\left| \nabla \phi(\mathbf x) \right|}
\end{align}
i.e. $\mathbf u$ must be parallel to the gradient of $\phi$ at $x_0$ and is guaranteed to be optimal in the limit of small $t$. Thus the problem becomes to find some $\epsilon$ that satisfies equation \ref{optimgoal} with
\begin{align} \label{gradupdate}
    \mathbf x_1 = \mathbf x_0 + \epsilon \nabla \phi
\end{align}

The computation of $\nabla \phi$ will be performed by the framework's autodiff algorithm. The computation of $\epsilon$ is the job of the Optimzer algorithm. With these, equation \ref{gradupdate} can be applied iteratively until a stationary point of $\phi$ is reached.

\subsection{Automatic Differentiation}
Consider the following function pseudocode
\begin{verbatim}
  def f(x):
      return x + cos(x)
\end{verbatim}
Its derivative can be written analytically as
\begin{verbatim}
  def df(x):
      return 1 - sin(x)
\end{verbatim}
Suppose instead that the function is
\begin{verbatim}
  def f(x):
      y = 0
      for k in range(len(x)):
          y += max(x, cos(y))
      y = concat(y, sqrt(abs(x)))
      return sum(y)
\end{verbatim}
Implementing the derivative of the above function is impractical. A possibility is to numerically approximate the partial derivatives using the definition.
\begin{verbatim}
  def df(x):
      h = 0.000000001
      dy = []
      for k in range(len(x)):
          x2 = x.copy()
          x2[k] += h
          dy.append((f(x2) - f(x)) / h)
      return dy
\end{verbatim}
This method has two issues:
\begin{itemize}
    \item It is prone to numerical instability and round-off errors (especially if working with low-precision floating point numbers).
    \item f(x) is evaluated $2|x|$ times. The computation cost becomes problematic when working with high dimensional functions.
\end{itemize}
Automatic differentiation, or autodiff for short, is a method that avoids both these issues. It makes use of the fact that any function is composed of more elementary functions for which the derivative is known analytically \cite{autodiff1981, autodiffpaper}. The chain rule of differentiation is used to assemble these to yield the total derivative.
\subsection{Partial Derivatives}
Before developing the autodiff algorithm, it is instrumental to understand the nuance of total derivatives, partial derivatives, and differentiating with variables held fixed. Let $y = f(a(x), b(x))$. The total derivative of $y$ with respect to $x$ is, by definition \cite{lhospital},
\begin{align}
    \frac{dy}{dx} = \lim_{h \to 0} \frac{f(a(x + h), b(x + h)) - f(a(x), b(x))}{h}
\end{align}
The partial derivative of $y$ with respect to $b$ is
\begin{align}
    \left(\frac{\partial y}{\partial b}\right)_a = \lim_{h \to 0} \frac{f(a(x), b(x) + h) - f(a(x), b(x))}{h}
\end{align}
where the subscript $a$ denotes the variable held fixed and in some cases is implied. However, its explicitness is necessary when partially differentiating with respect to $x$:
\begin{align}
    \left(\frac{\partial y}{\partial x}\right)_a = \lim_{h \to 0} \frac{f(a(x), b(x + h)) - f(a(x), b(x))}{h}
\end{align}

\subsection{The Chain Rule}
Let $y = f(a, b)$, $a = g(u, v)$, and $b = p(r, s)$ where $u, v, r, s$ are all functions of $x$. Using the chain rule, the total derivative of $y$ with respect to $x$ can be written in terms of partial derivatives as
\begin{align}
    \frac{dy}{dx} = \left(\frac{\partial y}{\partial a}\right)_b \frac{da}{dx} + \left(\frac{\partial y}{\partial b}\right)_a \frac{db}{dx}
\end{align}
The resulting total derivatives can be found by recursively applying the chain rule, e.g.
\begin{align}
        \frac{da}{dx} = \left(\frac{\partial a}{\partial u}\right)_v \frac{du}{dx} + \left(\frac{\partial a}{\partial v}\right)_u \frac{dv}{dx}
\end{align}
We then have
\begin{align}
    \frac{dy}{dx} &= \left(\frac{\partial y}{\partial a}\right)_b \left( \left(\frac{\partial a}{\partial u}\right)_v \frac{du}{dx} + \left(\frac{\partial a}{\partial v}\right)_u \frac{dv}{dx} \right) + \ldots
    \\
    &= \left(\frac{\partial y}{\partial a}\right)_b \left(\frac{\partial a}{\partial u}\right)_v \frac{du}{dx} + \left(\frac{\partial y}{\partial a}\right)_b \left(\frac{\partial a}{\partial v}\right)_u \frac{dv}{dx} + \ldots
    \\
    &= \left(\frac{\partial y}{\partial x}\right)_{b,v} + \left(\frac{\partial y}{\partial x}\right)_{b,u} +
    \left(\frac{\partial y}{\partial x}\right)_{a,r} +
    \left(\frac{\partial y}{\partial x}\right)_{a,s} \label{partials}
\end{align}
The pattern is now apparent: the total derivative is the summation of partial derivatives with different branches of the function tree held fixed. Although in this example the function tree is binary, it can be extended to any branching factor by induction.

Computing the terms in equation \ref{partials} will be performed sequentially, e.g.
\begin{align} \label{seq_deriv}
    \left(\frac{\partial y}{\partial a}\right)_{b} 
    \to \left(\frac{\partial y}{\partial u}\right)_{b,v} 
    \to \left(\frac{\partial y}{\partial x}\right)_{b,v} 
\end{align}
Thus, intermediary derivatives such as $(\partial a / \partial u)_v$ need not be computed. When the component functions are multivariate it is more trivial to compute $(\partial y / \partial u)_{b,v}$ given $(\partial y / \partial a)_b$ than computing $(\partial a / \partial u)_v$ and multiplying.

\subsection{Back Propagation}
To compute derivatives as in equation \ref{seq_deriv}, every elementary function 
\begin{align}
    f(a_1, a_2, \ldots a_n) = b
\end{align}
in the framework must implement the derivative function
\begin{align}
    f' \left( \frac{\partial y}{\partial b}, a_1, a_2, \ldots a_n \right) = \bigg\{ \frac{\partial y}{\partial a_1}, \frac{\partial y}{\partial a_2}, \ldots \frac{\partial y}{a_n} \bigg\}
\end{align}
where $y$ is the scalar function whose gradient we wish to compute. Suppose we construct the function
\begin{align}
    y &= f_2(b, x) \\
    b &= f_1(a, x)
\end{align}
where $x$ is the independent variable, $a$ is not a function of $x$, and $f_1, f_2$ are elementary functions implemented in the framework. We can compute $dy/dx$ as follows:
\begin{align}
    \bigg\{ \left(\frac{\partial y}{\partial b}\right)_x, \left(\frac{\partial y}{\partial x}\right)_b \bigg\} = f'_2(1, b, x)
\end{align}
using the fact that $\partial y / \partial y = 1$. Then
\begin{align} \label{weird}
    \bigg\{ \left(\frac{\partial y}{\partial x}\right)_{x,a}, \left(\frac{\partial y}{\partial b}\right)_{x, x} \bigg\} = f'_1\left( \left(\frac{\partial y}{\partial b}\right)_x, a, x \right)
\end{align}
The total derivative is the summation of the partials
\begin{align}
    \frac{dy}{dx} = \left(\frac{\partial y}{\partial x}\right)_{x,a} + \left(\frac{\partial y}{\partial x}\right)_b
\end{align}

It is now apparent why the name \say{back propagation} \cite{lecun1987phd} is suited - it starts with the output, computes the derivative with respect to the current variable, then recursively passes it back the computation tree to the variables that computed it until the independent variables (leaves) are reached.

Note: it may appear strange to hold $x$ fixed twice, or hold it fixed while differentiating with respect to it (in equation \ref{weird}). This is an important detail that must be understood. It may bring clarity to define $c = x$ and substitute it in $f_2$, then carry out the computation in a similar manner.

\section{Implementing Elementary Functions}
ArrayFlow implements the following elementary functions. This set is sufficient for building a large variety of popular network models and solving many other common optimization problems.
\begin{multicols}{2}
  \begin{itemize}
        \item matrix\_multiply(a, b)
        \item cross\_correlate(s, k)
        \item times(a, b)
        \item divide(a, b)
        \item max(a, b)
        \item min(a, b)
        \item maxpool(x, n)
        \item sum(x)
        \item add(a, b, c, \ldots)
        \item subtract(a, b)
        \item power(x, n)
        \item exponential(x)
        \item log(x)
        \item sqrt(x)
        \item sin(x)
        \item cos(x)
        \item tanh(x)
        \item mean(x)
        \item absolute\_value(x)
        \item concatenate(a, b, c, \ldots)
        \item expand(x)
        \item slice(x, start, end)
  \end{itemize}
\end{multicols}

\subsection{Einstein Notation}
When working with arrays, it is greatly convenient to use the Einstein indicial notation (also known as \say{the summation convention}) defined as follows: whenever an index is repeated exactly twice in a term, it implies summation over all values of that index. For example
\begin{align}
    A_{ij}b_i &= \sum_{i=1}^N A_{ij}b_i \\
    A_{ij}b_ic_j &= \sum_{i=1}^N \sum_{j=1}^M A_{ij}b_ic_j
\end{align}
Writing operations such as matrix multiplication or the inner product becomes quite compact:
\begin{align}
    AB = C \ \to \ C_{ij} = A_{ik}B_{kj} \\
    \langle a, b \rangle = a_ib_i
\end{align}
The Kronecker delta $\delta_{ij}$ is also useful \cite{chadwick}
\begin{equation}
 \delta_{ij} =
  \begin{cases}
    1 & \text{if i = j} \\
    0 & \text{otherwise}
  \end{cases}
\end{equation}

\subsection{Matrix Multiplication}
Let $A, B, C$ be matrices such that $C = AB$. Let $l$ be the scalar function which we are computing the gradient of (in optimization, this is usually a defined \say{loss} function, hence the letter $l$). Using Einstein notation and the chain rule
\begin{align}
    C_{rs} = A_{rp}B_{ps} \\
    \frac{\partial l}{\partial A_{ij}} = \frac{\partial l}{\partial} \frac{\partial C_{rs}}{\partial A_{ij}}
\end{align}
however, $\partial C_{rs} / \partial A_{ij} = 0$ unless $r=i$ and $p=j$, so we can make the substitution
\begin{align}
    \frac{\partial l}{\partial C_{is}}\frac{\partial}{\partial A_{ij}} A_{ij}B_{js} = \frac{\partial l}{\partial C_{is}B_{js}}
\end{align}
which reduces the double sum to a single sum. Although non-obvious, the last term can be written in regular matrix product notation as
\begin{align}
    \frac{\partial l}{\partial C_{is}}B_{js} = LB^T
\end{align}
where $L$ is defined as a shorthand
\begin{align}
    L_{ij} = \frac{\partial l}{\partial C_{ij}}
\end{align}
The derivative with respect to $B$ can be computed similarly
\begin{align}
    \frac{\partial l}{\partial B_{ij}} = \frac{\partial l}{\partial C_{rs}} \frac{\partial C_{rs}}{\partial B_{ij}} = \frac{\partial l}{\partial C_{rs}} \frac{\partial}{\partial B_{ij}}A_{rp}B_{ps}
\end{align}
Again, the last term is $0$ unless $p=i$ and $s=j$. Making the index substitution yields
\begin{align}
    \frac{\partial l}{\partial C_{rj}}A_{ri} = A^TL
\end{align}
\subsection{Max Pooling}
The maxpool function is not implemented in the numpy library. Here we'll implement the 1D case where the length of the array is divisible by the down-sampling factor $n$. Let 
\begin{equation}
 y = m(x, n)   
\end{equation}
denote the maxpool function. The array is divided into $k = \mathrm{len}(x) / n$ cells, and $y_k$ is the largest number in the $k$-th cell. Then, 

\begin{align}
 \frac{\partial l}{\partial x_i} =
  \begin{cases}
    \frac{\partial l}{\partial y_k} & \text{if $x_i$ is max of $k$-th cell} \\
    0 & \text{otherwise}
  \end{cases}
\end{align}
As an example, consider the array
\[ x = 
\begin{array}{|c|c|c|c|c|c|c|c|} \hline
    3 & 1 & -5 & 0 & 2 & 2 & 9 & 5 \\
    \hline
\end{array}
\]
The derivative is
\[
\frac{\partial l}{\partial x} =
\begin{array}{|c|c|c|c|c|c|c|c|} \hline
    \frac{\partial l}{\partial y_1} & 0 & 0 & \frac{\partial l}{\partial y_2} & \frac{\partial l}{\partial y_3} & 0 & \frac{\partial l}{\partial y_4} & 0 \\
    \hline
\end{array}
\]
This is implemented with the aid of few functions available in Numpy.
\subsection{Cross-Correlation (Convolution)} 
What most frameworks refer to as "convolution" is actually discrete cross-correlation. In the 1D case, it's equivalent to convolution with the kernel flipped \cite{dsp}. Let
\begin{equation}
c = \mathrm{ccor}(s, k)    
\end{equation}
denote the 1D cross-correlation of array $s$ (the \emph{signal}) with array $k$ (the \emph{kernel}). It is asserted that $\mathrm{len}(k) \le \mathrm{len}(s)$. Using Einstein notation, the operation can be written in indicial form as
\begin{align}
    c_i = k_js_{i+j}
\end{align}
Differentiating and making use of the Kronecker delta and its index substitution properties we have
\begin{align}
    \frac{\partial c_i}{\partial k_p} &= \delta_{pj} s_{i+j} \\
    \frac{\partial l}{\partial k_p} &= \frac{\partial l}{\partial c_i} \frac{\partial c_i}{\partial k_p} \\
    &= \frac{\partial l}{\partial c_i} \delta_{pj} s_{i+j} = \frac{\partial l}{\partial c_i} s_{i+p}
\end{align}
The last term can be written as another cross-correlation
\begin{align}
    \frac{\partial l}{\partial k} = \mathrm{ccor}\left(\frac{\partial l}{\partial c},  \ s \right)
\end{align}
Now derivative with respect to $s$
\begin{align}
    \frac{\partial c_i}{\partial s_n} &= k_j \delta_{(i+j)n} \\
    \frac{\partial l}{\partial s_n} &= \frac{\partial l}{\partial c_i} \frac{\partial c_i}{\partial s_n} \\
    &= \frac{\partial l}{\partial c_i} k_j \delta_{(i+j)n}
\end{align}
It's not obvious what the final term is; let us write out a few terms to understand the pattern.
\begin{align}
    k &= [k_0, k_1, k_2] \\
    s &= [s_0, s_1, \ldots, s_{20}] \\
    c &= [k_0s_0, k_1s_1, k_2s_2, \ldots, k_0s_{18}, k_1s_{19}, k_2s_{20}]
\end{align}
Now define, for an arbitrary variable $x$, $x'$ such that
\begin{equation}
    x_n' \coloneqq \frac{\partial l}{\partial x_n}
\end{equation}
Differentiating the above arrays we have
\begin{align}
    s_0' &= c_0'k_0 + 0k_1 + \ldots\\
    s_1' &= c_0'k_1 + c_1'k_0 \\
    s_{20}' &= c_{18}'k_2 + 0k_1 + \ldots
\end{align}
The pattern is now apparent and can be summarized in the following steps:
\begin{enumerate}
    \item Flip the kernel $k$.
    \item Append $\mathrm{len}(k) // 2$ zeroes to $c'$ (evenly on both sides).
    \item Cross-correlate $k$ with the modified $c'$.
\end{enumerate}
Remember that convolution in the 1D case is equivalent to cross-correlation with the kernel flipped. There is a convenient option in numpy's convolve function, called \say{mode}, which will zero pad the signal. Thus, the final derivative can be compactly computed by
\begin{equation}
    \frac{\partial l}{\partial s} = \mathrm{convolve} \left( k, \ \frac{\partial l}{\partial c}, \ \text{mode=\say{full}} \right)
\end{equation}

\subsection{Others}
The source code can be inspected for the implementation details of the other functions in ArrayFlow. Developed here is a subset of the more significant ones. 

For element-wise functions 
\begin{align}
    y &= f(x) \\
    y_i &= f(x_i)
\end{align}
$f'$ is given by the general form
\begin{equation}
    f'\left(\frac{\partial l}{\partial y}, \ x \right) = \frac{\partial l}{\partial y} * \frac{dy}{dx}
\end{equation}
where $*$ is the element-wise multiplication operator (numpy default).

\paragraph{Exponential}
The exponential function $y = e^x$ has derivative
\begin{equation}
    f' = \frac{\partial l}{\partial y} * e^x = \frac{\partial l}{\partial y} * y
\end{equation}
To avoid recomputing $f(x)$, we make $y$ a parameter of $f'$ in general:
\begin{equation}
    f'\left(\frac{\partial l}{\partial y}, \ x_1, x_2, \ldots, x_n \right) \to f'\left(\frac{\partial l}{\partial y}, y, \ x_1, x_2, \ldots, x_n \right)
\end{equation}
In most functions, $y$ is not needed and will be ignored, however it is useful in a few cases.

\paragraph{Square Root} 
The function $y = \sqrt x$ is another example where $y$ can save computation cost:
\begin{align}
    y^2 = x \\
    2y \frac{dy}{dx} = 1 \\
    f' = \frac{\partial l}{\partial y} * \frac{1}{2y}
\end{align}

\paragraph{Sum/Mean}
The summation function is quite simple to differentiate
\begin{align}
    y = \sum_i x_i \\
    \frac{dy}{dx_i} = 1
\end{align}
Therefore
\begin{align}
    f' = \frac{\partial l}{\partial y} * \mathbf{1}^{(n)}
\end{align}
where $\mathbf{1}^{(n)}$ is a length $n$ array of ones, in this case $n = \mathrm{len}(x)$ and the $*$ operation properly broadcasts as in numpy.
For the mean function, the derivative is identical up to the constant $1/n$.
\paragraph{Divide}
The element-wise division function $y = f(a, b) = a/b$ can be differentiated using the chain rule
\begin{align}
    \frac{\partial l}{\partial a} = \frac{\partial l}{\partial y} * \frac{1}{b} \\
    \frac{\partial l}{\partial b} = \frac{\partial l}{\partial y} * \frac{-a}{b^2} \\
    f'\left(\frac{\partial l}{\partial y}, y, a, b \right) = \bigg\{ \frac{\partial l}{\partial a}, \ \frac{\partial l}{\partial a}\bigg\}
\end{align}
where the division and square operations are also element-wise.

\paragraph{Min/Max}
The max function returns, element-wise, the greater of $a$ or $b$.
\begin{align}
    y = f(a, b) \\
    y_i = \mathrm{max}(a_i, b_i)
\end{align}
If $a_i > b_i$, the derivative $\partial l / \partial y_i$ is passed to $a_i$, and vice versa. Thus,
\begin{align}
 \frac{\partial l}{\partial a_i} =
  \begin{cases}
    \frac{\partial l}{\partial y_i} & \text{if $a_i > b_i$} \\
    0 & \text{otherwise}
  \end{cases}
\end{align}
and similarly for the derivative of $b_i$. For the $\mathrm{min}(a, b)$ function, simply flip the compare operators.

\paragraph{Absolute Value}
The absolute value function $y = |x|$ is not differentiable at $x=0$, however we can compute it for other points
\begin{align}
 \frac{\partial l}{\partial x_i} =
  \begin{cases}
    -\frac{\partial l}{\partial y_i} & \text{if $x_i<0$} \\
    \frac{\partial l}{\partial y_i} & \text{otherwise}
  \end{cases}
\end{align}
where we arbitrarily took the negative term for $x=0$. In practice, having an incorrect derivative (due to it being undefined) at a point is inconsequential because a variable taking on that value is improbable.

\paragraph{Concatenate}
Concatenating a list of $n$ arrays, where each array is of arbitrary length, requires mapping the derivative from the output (concatenated array) to the corresponding input. Thus, if the concat function is
\begin{align}
    y = f(x1, \ldots, x_n) \\
    \mathrm{len}(x_i) = m_i
\end{align}
the derivative of $l$ with respect to $x_i$ is found by slicing $\partial l/\partial y$ at the indices corresponding to the start and end of $x_i$ in $y$. These boundary points can be computed as the cumulative sum of $m_i$.
\paragraph{Slice}
Slicing keeps a contiguous subset of an array. The values not kept have zero derivative, while the ones kept have derivative equal to $\partial l / \partial y$.

\section{Code}
The framework must allow the user to define arbitrary scalar functions composed of the available set of elementary functions, and compute the derivative automatically. ArrayFlow implements this in an object oriented style consisting of Node objects and Operation subclasses.

\subsection{The Operation}
Operation is an abstract class that serves as a basis for the elementary functions available in ArrayFlow. Operation objects are never instantiated - instead, static methods for evaluating the function $f$ and its derivative, $f'$ are implemented. 

\begin{verbatim}
class Operation:
    
  @classmethod
  def evaluate(cls, *inputs) -> Node:
      inputs = list(inputs)
      for k, n in enumerate(inputs):
          if not isinstance(n, Node):
              inputs[k] = Constant(n)
      x = [n.array for n in inputs]
      y = cls._f(*x)
      output_node = Node(y)
      if is_training:
          output_node.op = cls
          output_node.input_nodes = inputs
      return output_node
  
  @classmethod
  def differentiate(cls, 
        node: Node) -> List[numpy.ndarray]:
      dldy = node.partial_derivative
      x = [n.array for n in node.input_nodes]
      y = node.array
      return cls._df(dldy, y, *x)
\end{verbatim}
The \texttt{evaluate()} method takes an arbitrary number of inputs of type Node and passes their arrays to \texttt{\_f(*x)} to compute $f$. The output is a Node whose \texttt{op} field is the operating class.

The \texttt{differentiate} method uses the parameter node's \texttt{dldy} = $\partial l / \partial y$ field to compute the derivatives with respect to the input nodes $\partial l / \partial x_i$.
\begin{verbatim}    
  @staticmethod
  def _f(*x: numpy.ndarray) -> numpy.ndarray:
      pass
  
  @staticmethod
  def _df(dldy: numpy.ndarray, y: numpy.ndarray, 
        *x: numpy.ndarray) -> List[numpy.ndarray]:
      pass
\end{verbatim}
The abstract static methods \texttt{\_f(*x)} and \texttt{\_df(dldy, y, *x)} are to be implemented by each elementary function in the framework.

\subsection{The Node}
Node is the superclass that represents all nodes in the computation tree.
\begin{verbatim}
class Node:

  op: 'Operation' = None
  input_nodes: List['Node'] = []
  array: numpy.ndarray = None
  _partial_derivative: numpy.ndarray = None
  _is_reset: bool = True
\end{verbatim}
The \texttt{op} field is a pointer to the Operation subclass which computed the current Node. Every variable's data is stored as a numpy ndarray in the \texttt{array} field. \texttt{\_partial\_derivative} is the shadow variable for the \texttt{partial\_derivative} property which stores $\partial l / \partial x$ where $x$ is the current node's variable array (and therefore has the same shape). We'll defer the discussion on the \texttt{\_is\_reset} field to the upcoming section on computing the gradient.

\begin{verbatim}
  def __init__(self, data):
      if not isinstance(data, numpy.ndarray):
          data = numpy.asarray(data, dtype=precision)
      if data.dtype != precision:
          data = data.astype(precision)
      self.array = data
\end{verbatim}
A Node object is instantiated by providing the variable's data, which is converted (if not already) to an ndarray of a selected precision (32-bit float by default).

\paragraph{Parameter and Constant}
There are two subclasses of Node. First is the Parameter, which represents the independent variables when computing the gradient, also known as \say{model parameters} or \say{trainable parameters} in ML lingo.

\begin{verbatim}
class Parameter(Node):

  def __init__(self, data):
      assert is_training
      super().__init__(data)
      self._partial_derivative = 
        numpy.zeros(self.array.shape, dtype=precision)

  @Node.partial_derivative.setter
  def partial_derivative(self, value: numpy.ndarray):
      self._is_reset = False
      if value.ndim == 1 + self._partial_derivative.ndim:
          value = numpy.sum(value, axis=-1)
      self._partial_derivative += value
\end{verbatim}

Parameter overrides the \texttt{partial\_derivative} property to implement the proper procedure for computing the total derivative. First, the \texttt{\_partial\_derivative} array must be instantiated with zeroes so that the partial derivatives (from different branches of the computation tree) can be added as per equation \ref{partials}. 

The array dimension check and subsequent sum over the last axis are there to handle the case in which a parameter is used in a computation that was broadcasted by numpy (usually when the Parameter is a single value multiplied by a vector such as time, as shown in one of the demos). The \texttt{\_is\_reset} member variable is set to False to mark that the derivatives began accumulation. This will be further discussed shortly.

The Constant class is the other subclass of Node, and, like Parameter, it represents a leaf in the computation tree. It therefore does not have input nodes or operation members.

\begin{verbatim}
class Constant(Node):

  @property
  def partial_derivative(self) -> None:
      return None

  @partial_derivative.setter
  def partial_derivative(self, value: numpy.ndarray):
      pass
\end{verbatim}
Further, since this node represents constants, it ignores the derivatives - autodiff still computes them but they simply aren't stored when passed to a Constant node.

\subsection{Computing The Gradient}
The scalar field whose gradient we are computing (most commonly the \say{loss}) will be a Node, so the gradient computation function is implemented in the Node class. 
\begin{verbatim}
  def _reset_parameter_derivatives(self):
      if not self._is_reset:
          self._partial_derivative.fill(0)
          self._is_reset = True
      for node in self.input_nodes:
          node._reset_parameter_derivatives()
\end{verbatim}
First, the partial derivatives of all Parameter objects must be reset to zeroes to ready them for accumulation. Since this step is unnecessary for intermediary nodes and is thus skipped, the \texttt{\_is\_reset} boolean member is used to indicate when resetting is appropriate.

\begin{verbatim}
  def compute_gradient(self):
      assert self.array.size == 1
      assert is_training
      if self.partial_derivative is None:
          self.partial_derivative = 
              numpy.ones_like(self.array)
      self._reset_parameter_derivatives()
      self._autodiff()

  def _autodiff(self):
      if self.op is not None:
          dldx = self.op.differentiate(self)
          for k, pd in enumerate(dldx):
              self.input_nodes[k].partial_derivative = pd
      for node in self.input_nodes:
          node._autodiff()
\end{verbatim}

The recursive computation of the gradient is kicked off by calling the \texttt{compute\_gradient()} method on the respective node. It is enforced that the node is a scalar field (size of the data array = 1). The \texttt{is\_training} global variable must be set to True for differentiation to be available. If False, ArrayFlow is in inference mode and will not allocate memory for the derivatives.

The function then starts the autodiff back propagation process at the top with $\partial l/\partial l = 1$, passing it to the input nodes, and so on down the computation tree until the leaves (Parameter or Constant nodes) are reached.

\section{The Optimizer}
Recall from equation \ref{gradupdate} that given a point $\mathbf x_0$, another point $\mathbf x_1$ can be found such that $f(\mathbf x_1) < f(\mathbf x_0)$ by adding a small vector in the direction antiparallel to the gradient. Repeating this process will get arbitrarily close to a locally optimum point.

\paragraph{Inertia}
Anticipating noise due to the chaotic nature of differentiation \cite{momentum}, we shall limit the contribution of a single step's gradient to the stepping direction by using the exponential moving average of the direction vector $\mathbf{g}$ instead:
\begin{align}
    \mathbf{g}_n = \beta \nabla l(\mathbf x_n) + (1 - \beta) \mathbf{g}_{n-1} \\
    \mathbf x_{n + 1} = \mathbf x_n - \epsilon \mathbf{g}_n
\end{align}
where $\beta$ is the weigth coefficient.

The necessity of having proper values for $\epsilon$ and $\mathbf x_0$ becomes apparent. Finding a good starting point $\mathbf{x_0}$ is important \cite{init}, as is shown in one of the accompanying demos, however for most problems there is little to no theory on how to solve for it in advance. Sometimes domain-specific knowledge can be used to make an educated guess, while other times it's chosen randomly.

Solutions for the value of $\epsilon$, also known as the \say{step size} can be methodically computed. The optimizer in ArrayFlow uses heuristics to change $\epsilon$ between gradient update iterations (steps), treating it an optimization problem inside the main optimization problem.

\paragraph{Stepping Heuristic}
The $\epsilon$ optimization algorithm is based on the following rules, derived from empirical observations during experimentation. Starting with a small initial guess $\epsilon_0 << 1$ and given the loss as a function of steps
\begin{itemize}
    \item $\epsilon$ is slowly increased if the loss is decreasing monotonically.
    \item When the loss first increases, $\epsilon$ can never again increase.
    \item $\epsilon$ can decrease if the loss is increasing and concave up.
\end{itemize}
The optimizer is detailed using pseudocode in Algorithm \ref{algo:optimizer}.

\begin{algorithm}
\caption{Optimizer} \label{algo:optimizer}
\KwIn{m, $\beta$, $s_0$, $x_0$}
$s \gets s_0$ \\
$x \gets x_0$ \\
$r \gets \texttt{true}$ \\ 
$l_t \gets \texttt{new queue}(l_0)$ \\
$\overline{l_t} \gets \texttt{new queue}([l_t] * 4)$ \\
$c_1 \gets [-1/3, 3/2, -3, 11/6]$ \\
$c_2 \gets [-1, 4, -5, 2]$ \\
$\nabla f_{t-1} \gets 0$ \\
\Repeat{\upshape $x$ converges} {
$l \gets f(x)$ \\
\upshape \texttt{push} $l$ onto $l_t$ \\
\If{\upshape size($l_t$) $>$ m} {
\upshape \texttt{remove} from $l_t$ \\
}
\upshape \texttt{push} mean($l_t$) onto $\overline{l_t}$ \\
\upshape \texttt{remove} from $\overline {l_t}$ \\

$\frac{d\overline{l_t}}{dt} \gets c_1 * \overline{l_t}$ \\
$\frac{d^2\overline{l_t}}{dt^2} \gets c_2 * \overline{l_t}$ \\

\uIf{$\frac{d\overline{l_t}}{dt} > 0$ \upshape and $\frac{d^2\overline{l_t}}{dt^2} > 0$} {
$s \gets 0.99s$ \\
    \If{$r$} {
    $r \gets \texttt{false}$
    }
}
\ElseIf{$r$ \upshape and $\frac{d\overline{l_t}}{dt} < 0$}{
$s \gets 1.02s$ \\
}
$\nabla f_t \gets \texttt{autodiff}(l)$ \\
$x \gets \beta \nabla f_t + (1 - \beta) \nabla f_{t-1}$ \\
$\nabla f_{t-1} \gets \nabla f_t$ \\
}
\Return{x}
\end{algorithm}

\section{Demos}
Three problems are solved using ArrayFlow as demonstrations. The mathematics of the problems are set up here, while the code and results are in the accompanying Demos notebook on GitHub.
\subsection{The Catenary}

An interesting problem is that of finding the shape of a hanging rope fixed at two points in a uniform gravitational field. This curve is called a \emph{catenary}. The rope is modeled as having infinite stiffness, uniform mass density, and being in a state of pure tension with zero moment everywhere.

\begin{figure}
  \centering
  \includegraphics[width=.9\linewidth]{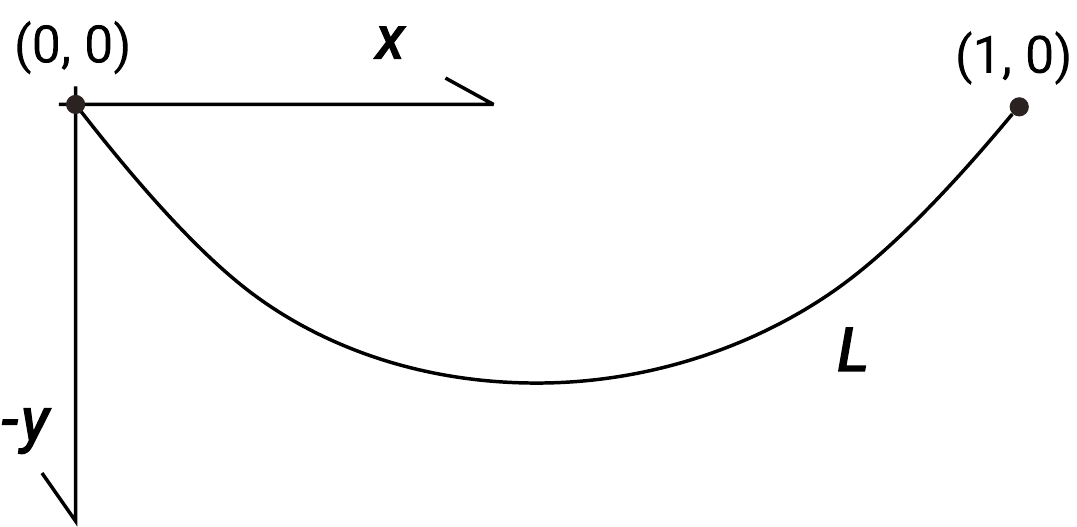}
  \caption{Coordinate system for the catenary problem.}
  \Description{Coordinate system for the catenary problem.}
  \label{fig:catenarydiagram}
\end{figure}

The coordinate system is set up as in figure \ref{fig:catenarydiagram}, where the rope length $L > 1$. The rope is discretized into $n$ segments

\begin{align}
    L = \int_c d\vec{s} \approx \sum_i \sqrt{\Delta x_i^2 + \Delta y_i^2} \\
    \Delta x_i = 1/n
\end{align}
Observe that the vector $\Delta y$ can be written using the cross-correlation of $y$ with kernel $[1, -1]$
\begin{align}
    \Delta y &= [y_0 - y_1, \ldots, y_{n-1} - y_n] \\
    &= \mathrm{ccor}(y, [1, -1])
\end{align}
Now let
\begin{align}
    s = \sqrt{1/n + \Delta y^2}
\end{align}
where all operations are element-wise. It follows that
\begin{align}
    L = \sum s_i
\end{align}
At equilibrium, the rope will take on an arrangement that minimizes its potential energy $E$ where
\begin{align}
    E = \int_c y \ d \vec s
\end{align}
$E$ can be approximated by finding the height of the centroid of each discrete segment times its mass
\begin{align}
    E \approx \langle s, \mathrm{ccor}(y, [1/2, 1/2]) \rangle
\end{align}
This defines a constrained optimization problem: minimize $E$ with constraints
\begin{align}
    L = L_0 \\
    y_0 = 0 \\
    y_n = 0
\end{align}
To solve this problem numerically, we initialize $y$ randomly and set the end points to $0$. We define a loss function
\begin{align}
    l(y) = (L - L_0)^2 + E
\end{align}
This will create a contention between the lowering of the rope's center of mass and its length equaling $L_0$. The rope could lower if allowed to stretch, which is a different problem called the \say{elastic catenary}. 

Although not included in the loss function, the fixed endpoint conditions must be satisfied. To this end, a \say{trick} is employed as so: first, compute the gradient
\begin{align}
    (\nabla l)_i = \frac{dl}{dy_i}
\end{align}
then set
\begin{align}
    \frac{dl}{dy_0} = 0 \\
    \frac{dl}{dy_n} = 0
\end{align}
before performing the gradient step on $y$. This will ensure that the endpoints will never move from their original position.

\begin{figure}
  \centering
  \includegraphics[width=1.04 \linewidth]{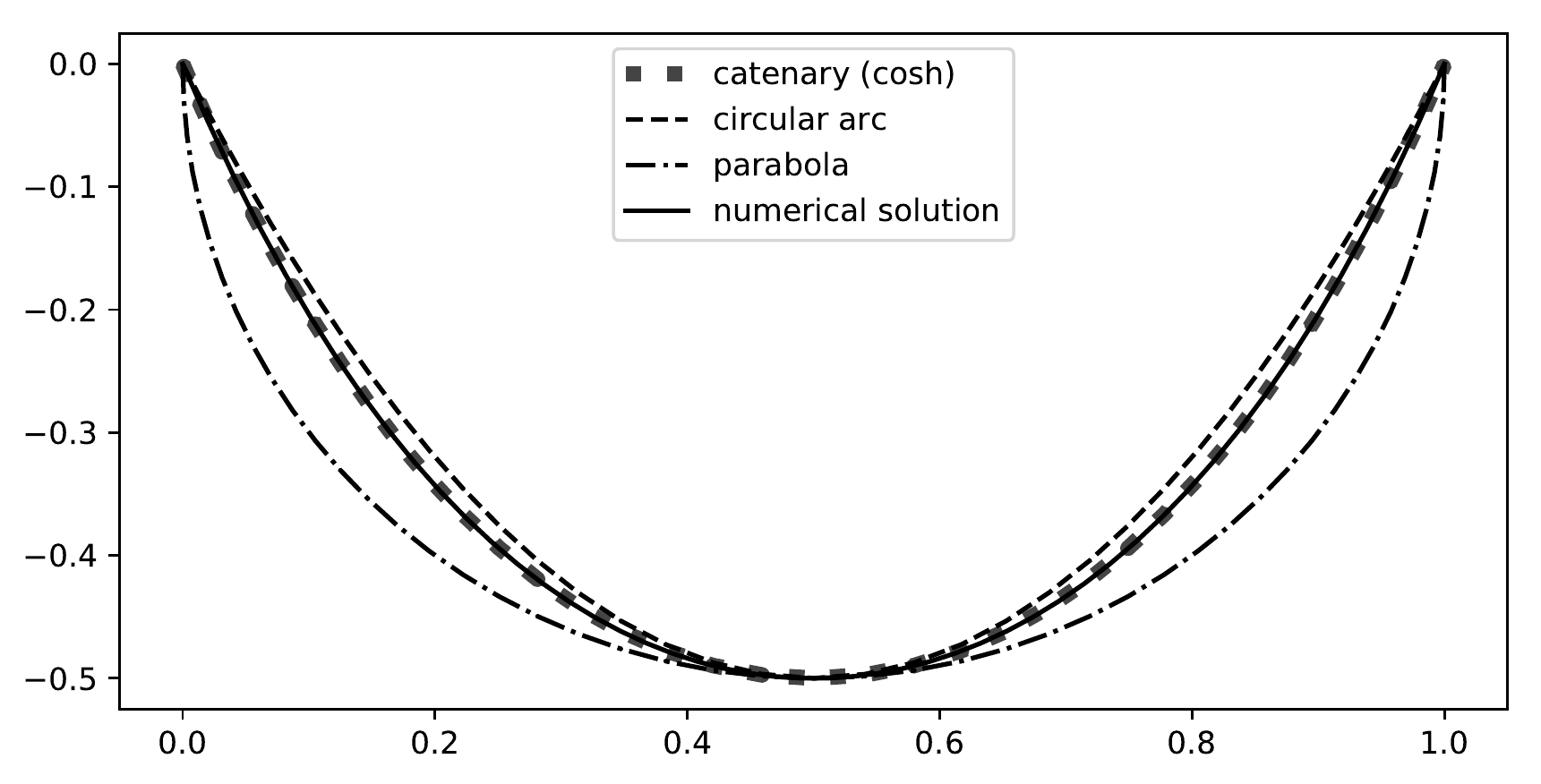}
  \caption{Plot of the numerical solution and other curves.}
  \Description{Coordinate system for the catenary problem.}
  \label{fig:catplot}
\end{figure}

The algorithm converged $y$ to the curve in figure \ref{fig:catplot}. The catenary $y_c$ can be solved analytically using the Calculus of Variations for the general solution which is the hyperbolic cosine curve 
\begin{align}
y_c = a\cosh\left(\frac{x - b}{a}\right) + c
\end{align}

Several other curves are plotted in figure \ref{fig:catplot}: a circular arc $y_r$, a parabola $y_p$, the catenary $y_c$, and the numerical solution $y$. All these curves pass through the 3 points $(0,0), (1, 0), (1/2, -1/2)$.
\begin{align}
    r^2 = (x - x_0)^2 + (y_c - y_0)^2 \\
    y_p = c_1x + c_2x + c_3 \\
    x \in [0, 1] \\
    y_i(0) = 0 \\
    y_i(1) = 1 \\
    y_i(1/2) = -1/2
\end{align}
Since $y_p$, $y_c$, and $y_r$ have 3 parameters, there exists a unique solution for each:
\begin{align}
    y_c = 0.3094 \cosh\left(\frac{x - 0.5}{0.3094}\right) - 0.8094 \\
    y_r = -\sqrt{0.25 - (x - 0.5)^2} \\
    y_p = 2x(x-1)
\end{align}
As seen in figure \ref{fig:catplot} the optimizer found a very precise numerical approximation (solid line) to the catenary (dotted line).

\subsection{Classifying Histograms}
\begin{figure}
  \centering
  \includegraphics[width=.9\linewidth]{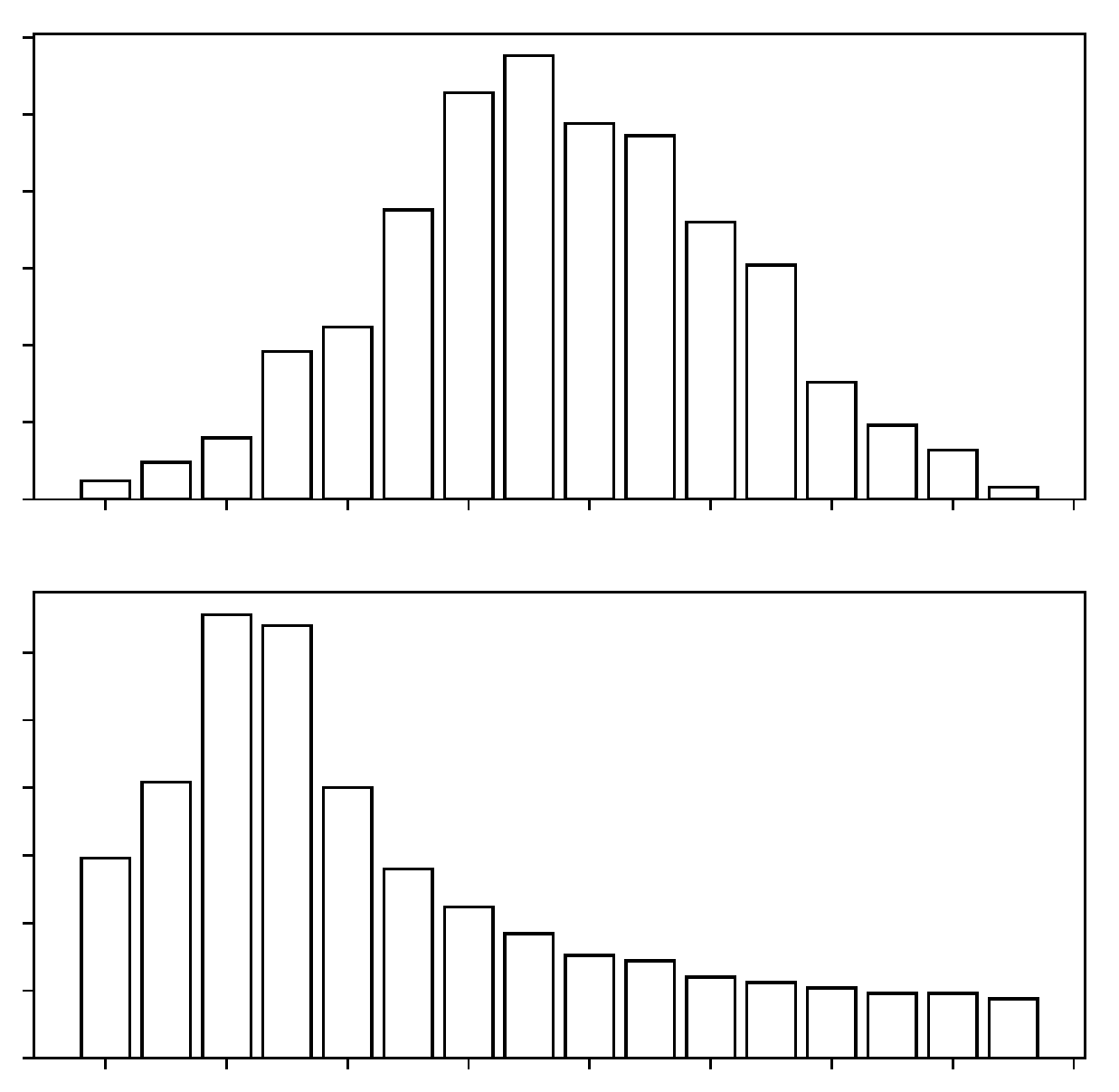}
  \caption{Example of $x$ from class 1 (top) and class 0 (bottom).}
  \Description{Example of $x$ from class 1 (top) and class 0 (bottom).}
  \label{fig:histclasses}
\end{figure}

With a quick glance, a human can accurately resolve whether a histogram likely represents data from a Gaussian distribution. This vision problem is posed as follows: draw 500 points from either $\mathcal N(0, 1)$ (class 1, figure \ref{fig:histclasses} top) or $\mathcal D$ (class 0, figure \ref{fig:histclasses} bottom) where $\mathcal{D}$ is some other distribution. Then compute the histogram with 16 bins, i.e. the counts for each bin given by $h_i, \ i=1, 2, \ldots, 16$. Let
\begin{align}
    x: x_i = \frac{h_i}{\sum h_i}
\end{align}
denote the normalized heights of the 16 bins, i.e. the fractions. This is the input to the classifier.

By fitting example data, a function will be learned $f: x \to \mathbb{R}$ where $f(x) > 0$ corresponds to class 1. The function (model), a small 1D \say{convolutional} network, will be constructed as so:

\begin{align}
    c_i &= \mathrm{ccor}(x, k_i) \ , \ \ i = 1, 2, 3 \\
    p_i &= \mathrm{maxpool}(c_i, 2) \\
    v &= \mathrm{concat}(p_1, p_2, p_3) \\
    y_0 &= W_1v + w_1 \\
    y &= \mathrm{sum}(y_0 \cdot w_2)
\end{align}
where model parameters and dimensions are as follows:
\begin{align}
    k_i&: (5 \times 1) \\
    W_1&: (7 \times 18) \\
    w_1, w_2&: (7 \times 1)
\end{align}
These parameters are initialized with random numbers drawn from $N(0, 1)$ and computed to optimize the loss function discussed next.

A common mapping of $\mathbb{R} \to (0, 1)$ is the logistic function
\begin{align}
    \sigma(x) = \frac{1}{1 + e^{-x}}
\end{align}
If the output $\hat y$ is constrained to $(0, 1)$ a good loss function is the cross-entropy
\begin{align}
    l(\hat y, y_t) = -y_t \log(\hat y) - (1 - y_t)\log(1 - \hat y) 
\end{align}
Computation cost can be saved by directly using $y$ and simplifying to
\begin{align}
l(y, y_t) &= -y_t \log(\sigma(y)) - (1 - y_t)\log(1 - \sigma(y)) \\
 &= \log(1 + e^y) - y_ty
\end{align}

Training the model was performed by creating batches of 20 random examples from each class and taking the mean of $l$ across a batch. After 1-2 minutes of optimization, the model achieves >99\% accuracy as this is not a particularly difficult problem. 

\paragraph{Hacking The Model}
\begin{figure}
  \centering
  \includegraphics[width=\linewidth]{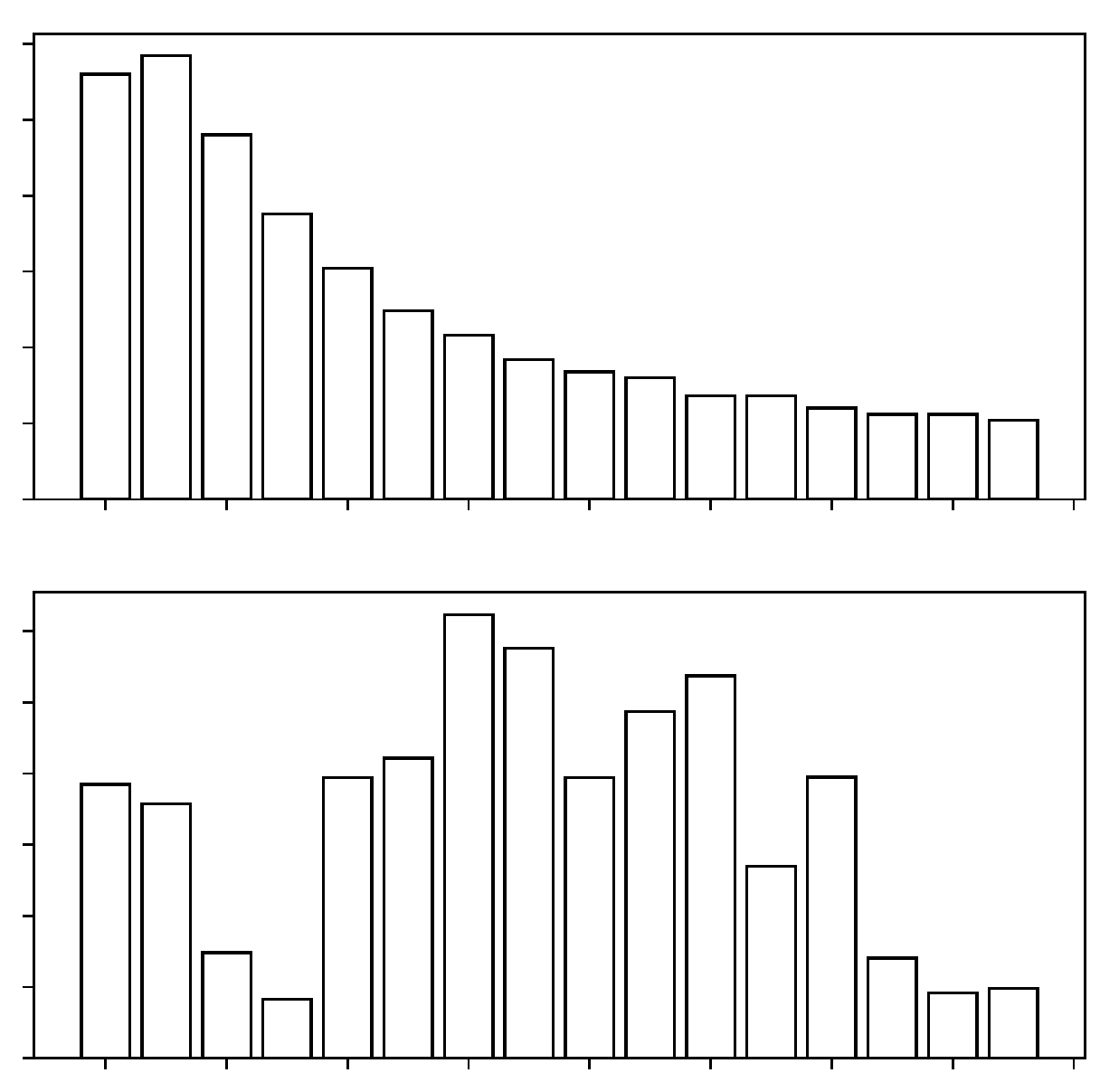}
  \caption{Example of $x$ from class 0 (top) morphed into class 1 by following the gradient (bottom).}
  \Description{Example of $x$ from class 0 (top) morphed into class 1 by following the gradient (bottom).}
  \label{fig:morph}
\end{figure}
We can make $x$ a Parameter object and compute the gradient of $l$ with respect to $x$. Given a data point $x_0$ that evaluates to class 0, i.e. $f(x_0) < 0$ we can compute
\begin{align}
    x_1 = x_0 + \epsilon \nabla f(x_0)
\end{align}
which has the property, for some sufficiently small $\epsilon > 0$
\begin{align}
    f(x_1) > f(x_0)
\end{align}
We can repeat this computation on $x_1$ with the new gradient $\nabla l = \nabla f(x_1)$ and iterate
until $f(x_1) > 0$ meaning that the original input $x_0$ was morphed into class 1. 

Figure \ref{fig:morph} top shows an example $x_0$ with $f(x_0) = -4.62$ (class 0) that has been changed in this manner to produce $x_1$ with $f(x_1) = 2.85$ (class 1) after several update iterations, shown in figure \ref{fig:morph} bottom. This is the concept behind adversarial networks \cite{hack}.

\subsection{Differential Equations}
ArrayFlow can be used to numerically solve problems involving differential equations. Consider the following ordinary differential equation (ODE) with boundary conditions
\begin{align} 
    2 \frac{d^2y}{dt^2} + \frac{dy}{dt} + 2y = 0  \label{ODE} \\
    y(0) = 1, \ \frac{dy}{dt} \bigg \rvert _0 = 1 \label{InitialConditions}
\end{align}
This \nth{2} order linear equation belongs to one of the most popular classes of ODEs as it models a damped oscillator and has many applications in physics and engineering. The analytical solution can be found by solving a characteristic polynomial, using the Laplace transform, or other methods; it is
\begin{align}
    y(t) = \frac{1}{3} e^{-t/4}\left(\sqrt{15} \sin(\sqrt{15} t / 4) + 3 \cos(\sqrt{15} t / 4)\right)
\end{align}
Numerically approximating the solution to the initial value problem is trivial using time-stepping algorithms such as Euler's method. For example, let
\begin{align}
    y_1 = y \\
    y_2 = \frac{dy}{dt}
\end{align}
Now substitute into equation \ref{ODE} to get, in matrix form,
\begin{align}
    \frac{d}{dt} \begin{bmatrix} y_1 \\ y_2 \end{bmatrix} 
    = \begin{bmatrix} 0 & 1 \\ -1 & -\frac{1}{2} \end{bmatrix}
    \begin{bmatrix} y_1 \\ y_2 \end{bmatrix}
\end{align}
We can then step through the approximate solution using
\begin{align}
    \begin{bmatrix} y_1 \\ y_2 \end{bmatrix}_t
    = \begin{bmatrix} y_1 \\ y_2 \end{bmatrix}_{t-1}
    + \Delta t \frac{d}{dt} \begin{bmatrix} y_1 \\ y_2 \end{bmatrix}_{t-1}
\end{align}
Let us increase the difficulty and solve the boundary value problem instead, i.e. given 
\begin{align}
    y(0) = 1, \ y(t_1) = b
\end{align}
The approach to approximating the solution using ArrayFlow is as follows:
\begin{enumerate}
    \item Discretize $y(t)$ into $N$ points in the range $[0, t_1]$.
    \item Initialize $y$ with a curve that satisfies the boundary conditions.
    \item Numerically compute the first and second derivatives at all $N$ points.
    \item Define the loss as the square of the LHS of equation \ref{ODE}.
    \item Optimize.
\end{enumerate}
The derivatives are computed using finite difference. The center difference formulas are \cite{kutz}
\begin{align}
    \frac{dy_n}{dt} = \frac{\frac{1}{12} y_{n-2} - \frac{2}{3} y_{n-1} + \frac{2}{3} y_{n+1} - \frac{1}{12} y_{n+2}}{\Delta t} + \mathcal O(\Delta t^4)
\\
    \frac{d^2y_n}{dt^2} = \frac{-\frac{1}{12} y_{n-2} - \frac{4}{3} y_{n-1} - \frac{5}{2} y_{n} + \frac{4}{3} y_{n+1} - \frac{1}{12} y_{n+2}}{\Delta t^2} + \mathcal O(\Delta t^4)
\end{align}
where in our case
\begin{align}
\Delta t = t_1 / (N - 1)
\end{align}
For the first two points  $n = 1, 2$ the center difference formulas above cannot be used so the forward formulas must substitute:
\begin{align}
    \frac{dy_n}{dt} = \frac{-\frac{25}{12} y_{n} + 4 y_{n+1} - 3 y_{n+2} + \frac{4}{3} y_{n+3} - \frac{1}{4} y_{n+4} }{\Delta t} + \mathcal O(\Delta t^4)
\\
    \frac{d^2y_n}{dt^2} = \frac{\frac{35}{12} y_{n} - \frac{26}{3} y_{n+1} + \frac{19}{2} y_{n+2} - \frac{14}{3} y_{n+3} + \frac{11}{12} y_{n+4}}{\Delta t^2} + \mathcal O(\Delta t^3)
\end{align}

The last two points in $y$ likewise must use the backward difference formulas. The coefficients for the \nth{1} and \nth{2} derivatives are the negative and same of those in the forward formulas, respectively.

During training, the boundary value constraints are imposed by setting $y_1 = 1, y_n = 0.1$ at every step. The derivative operator amplifies spikes due to noise or large changes in $y_m$ to $y_{m+1}$. For this reason, $y$ must be initialized with a curve that smoothly connects the boundary values, and optimization must be performed in small steps to avoid numerical instability. 
\begin{figure}
  \centering
  \includegraphics[width=\linewidth]{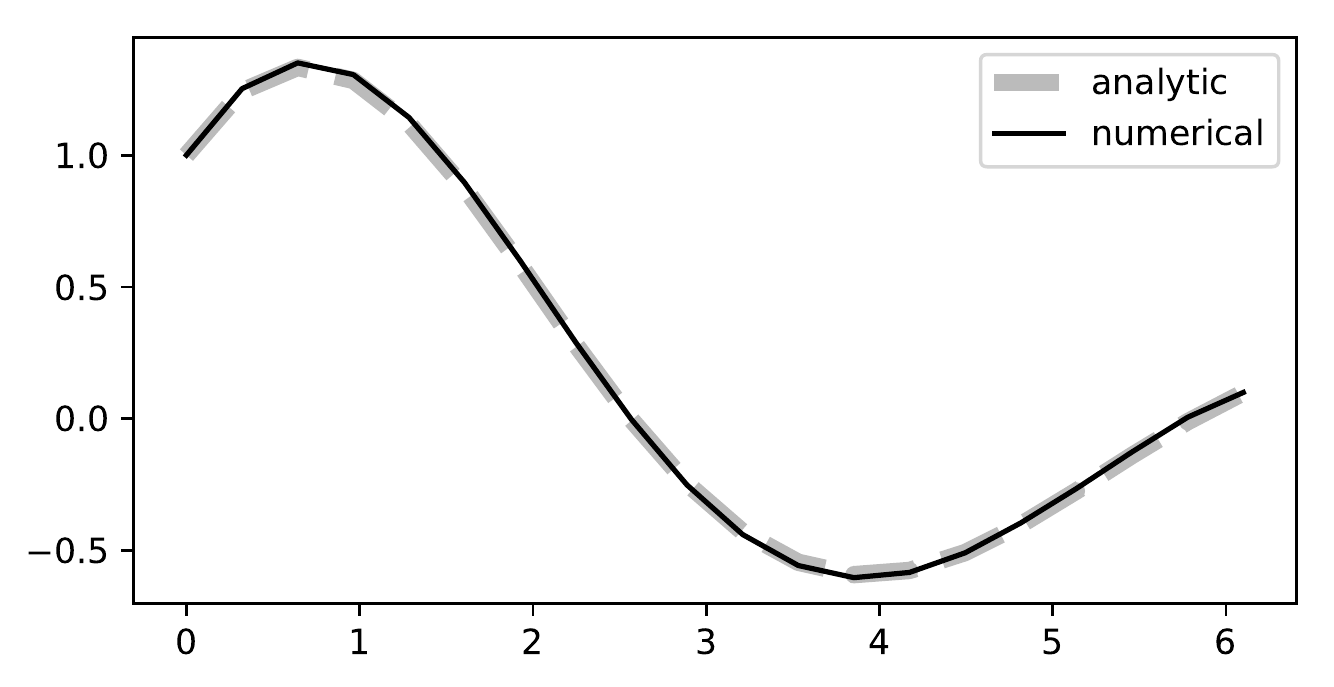}
  \caption{Solution of the ODE in equation \ref{fig:ode}}
  \Description{Solution of the ODE in equation \ref{fig:ode}}
  \label{fig:ode}
\end{figure}

If using 32-bit precision (ArrayFlow default) the upper bound of $N$ is roughly 20. Using a smaller step size will cause numerical overflow due to the $\Delta t^2$ term; for $N>20$ the floating point precision must be increased.

After optimization, $y$ converged to the curve in figure \ref{fig:ode}, a close match to the analytic solution. Although this is quite an inefficient method of solving ODEs, the exercise demonstrates the capabilities of ArrayFlow in this domain.

\section{Remarks}
Having completed the fully functional deep learning framework, below is a discussion pertaining to miscellaneous details.
\subsection{Static vs. Dynamic Computation Graphs} 
In ArrayFlow, each time a model function is evaluated, new Node and Constant objects are created. This leads to a high degree of flexibility - for example the entire model can be changed partway through optimization while still using the same Parameter objects. While this may have implications such as high discontinuity of gradient between steps, the framework is fully capable of handling changes without issue, as every model evaluation is a new computation tree.

On instantiation of a Node (or subclass) object, new memory space is requested for the data array. This could be avoided if the objects for a particular function's computation tree are pre-allocated and only the values of their data arrays modified in each subsequent call, increasing performance but foregoing the aforementioned flexibility.

\subsection{Distributed Training} Optimizing a function on multiple machines can be performed in the following two ways.

\paragraph{Splitting the Data} Let $X = \{x_1, x_2, \ldots, x_m\}$ denote a training data set. We'll create smaller disjoint subsets $X_i \subset X$ which need not be equal in cardinality. In general
\begin{align}
    X_i \cap X_j = \varnothing \\
    X_1 \cup X_2 \cup \ldots X_m = X
\end{align}
If all training examples $x_k \in X$ are assigned equal weight, the total loss is the mean loss across each subset
\begin{align}
    l = \frac{1}{|X|} \sum_k f(x_k, \mathbf p) = \sum_i \left(  \frac{1}{|X_i|}\sum_{x_j \in X_i} f(x_j, \mathbf p) \right)
\end{align}
Where $\mathbf p$ is the vector of parameters to be learned. It follows that
\begin{align}
    \nabla l = \frac{\sum_i|X_i| \nabla f(X_i, \mathbf p)}{\sum_i |X_i|}
\end{align}
where for each subset $X_i$, $f(X_i, \mathbf p)$ and $\nabla f(X_i, \mathbf p)$ can be computed by a different machine. After the gradient step, however, all machines must get the updated parameters $\mathbf p_{t+1}$
which means the time per step is that of the slowest machine. The tasks' time can be balanced by giving slower machines less data, i.e. making $|X_i|$ proportional to the compute power of machine $M_i$.

\paragraph{Splitting the Computation Tree}
The computation tree of a composite function in ArrayFlow can be partitioned into sub-trees where each sub-tree can be placed on a different machine. Consider a Node $n_i$ whose inputs are $n_1, n_2$ and data array is 
\begin{align}
   n_i.\texttt{arr} = y_i
\end{align}
For the forward pass, $y_1$ and $y_2$ are passed to $n_i$ which computes 
\begin{align}
    y_i = f_i(y_1, y_2)
\end{align}
on machine $M_i$. However, $y_1$ and $y_2$ need not be computed on $M_i$.

In the autodiff backward pass, $\partial l/\partial y_i$ is passed to $n_i$ which is to compute
\begin{align}
    \frac{\partial l}{\partial y_1}, \frac{\partial l}{\partial y_2} = df_i \left( \frac{\partial l}{\partial y_i}, y_i, y_1, y_2 \right)
\end{align}
on $M_i$. These arrays are passed to the host machines of $n_1$ and  $n_2$ respectively. Thus, an orchestrator that manages the mapping and interfaces between nodes and machines can achieve distributed inference and autodiff. The caveat is that data transfer between machines can be substantial, therefore this is maximally effective when the computation load for a sub-tree of Nodes greatly exceeds the data transfer latency.

\bibliographystyle{ACM-Reference-Format}
\nocite{*}
\bibliography{refs}


\begin{thebibliography}{14}


\ifx \showCODEN    \undefined \def \showCODEN     #1{\unskip}     \fi
\ifx \showDOI      \undefined \def \showDOI       #1{#1}\fi
\ifx \showISBNx    \undefined \def \showISBNx     #1{\unskip}     \fi
\ifx \showISBNxiii \undefined \def \showISBNxiii  #1{\unskip}     \fi
\ifx \showISSN     \undefined \def \showISSN      #1{\unskip}     \fi
\ifx \showLCCN     \undefined \def \showLCCN      #1{\unskip}     \fi
\ifx \shownote     \undefined \def \shownote      #1{#1}          \fi
\ifx \showarticletitle \undefined \def \showarticletitle #1{#1}   \fi
\ifx \showURL      \undefined \def \showURL       {\relax}        \fi
\providecommand\bibfield[2]{#2}
\providecommand\bibinfo[2]{#2}
\providecommand\natexlab[1]{#1}
\providecommand\showeprint[2][]{arXiv:#2}

\bibitem[\protect\citeauthoryear{{Abadi}, {Agarwal}, {Barham}, {Brevdo},
  {Chen}, {Citro}, {Corrado}, {Davis}, {Dean}, {Devin}, {Ghemawat},
  {Goodfellow}, {Harp}, {Irving}, {Isard}, {Jia}, {Jozefowicz}, {Kaiser},
  {Kudlur}, {Levenberg}, {Mane}, {Monga}, {Moore}, {Murray}, {Olah},
  {Schuster}, {Shlens}, {Steiner}, {Sutskever}, {Talwar}, {Tucker},
  {Vanhoucke}, {Vasudevan}, {Viegas}, {Vinyals}, {Warden}, {Wattenberg},
  {Wicke}, {Yu}, and {Zheng}}{{Abadi} et~al\mbox{.}}{2016}]%
        {tensorflow}
\bibfield{author}{\bibinfo{person}{Mart{\'\i}n {Abadi}},
  \bibinfo{person}{Ashish {Agarwal}}, \bibinfo{person}{Paul {Barham}},
  \bibinfo{person}{Eugene {Brevdo}}, \bibinfo{person}{Zhifeng {Chen}},
  \bibinfo{person}{Craig {Citro}}, \bibinfo{person}{Greg~S. {Corrado}},
  \bibinfo{person}{Andy {Davis}}, \bibinfo{person}{Jeffrey {Dean}},
  \bibinfo{person}{Matthieu {Devin}}, \bibinfo{person}{Sanjay {Ghemawat}},
  \bibinfo{person}{Ian {Goodfellow}}, \bibinfo{person}{Andrew {Harp}},
  \bibinfo{person}{Geoffrey {Irving}}, \bibinfo{person}{Michael {Isard}},
  \bibinfo{person}{Yangqing {Jia}}, \bibinfo{person}{Rafal {Jozefowicz}},
  \bibinfo{person}{Lukasz {Kaiser}}, \bibinfo{person}{Manjunath {Kudlur}},
  \bibinfo{person}{Josh {Levenberg}}, \bibinfo{person}{Dan {Mane}},
  \bibinfo{person}{Rajat {Monga}}, \bibinfo{person}{Sherry {Moore}},
  \bibinfo{person}{Derek {Murray}}, \bibinfo{person}{Chris {Olah}},
  \bibinfo{person}{Mike {Schuster}}, \bibinfo{person}{Jonathon {Shlens}},
  \bibinfo{person}{Benoit {Steiner}}, \bibinfo{person}{Ilya {Sutskever}},
  \bibinfo{person}{Kunal {Talwar}}, \bibinfo{person}{Paul {Tucker}},
  \bibinfo{person}{Vincent {Vanhoucke}}, \bibinfo{person}{Vijay {Vasudevan}},
  \bibinfo{person}{Fernanda {Viegas}}, \bibinfo{person}{Oriol {Vinyals}},
  \bibinfo{person}{Pete {Warden}}, \bibinfo{person}{Martin {Wattenberg}},
  \bibinfo{person}{Martin {Wicke}}, \bibinfo{person}{Yuan {Yu}}, {and}
  \bibinfo{person}{Xiaoqiang {Zheng}}.} \bibinfo{year}{2016}\natexlab{}.
\newblock \showarticletitle{{TensorFlow: Large-Scale Machine Learning on
  Heterogeneous Distributed Systems}}.
\newblock \bibinfo{journal}{\emph{arXiv e-prints}}, Article
  \bibinfo{articleno}{arXiv:1603.04467} (\bibinfo{date}{March}
  \bibinfo{year}{2016}), \bibinfo{numpages}{arXiv:1603.04467}~pages.
\newblock
\showeprint[arxiv]{1603.04467}~[cs.DC]


\bibitem[\protect\citeauthoryear{Baydin, Pearlmutter, Radul, and
  Siskind}{Baydin et~al\mbox{.}}{2017}]%
        {autodiffpaper}
\bibfield{author}{\bibinfo{person}{At\i{}l\i{}m~G\"{u}nes Baydin},
  \bibinfo{person}{Barak~A. Pearlmutter}, \bibinfo{person}{Alexey~Andreyevich
  Radul}, {and} \bibinfo{person}{Jeffrey~Mark Siskind}.}
  \bibinfo{year}{2017}\natexlab{}.
\newblock \showarticletitle{Automatic Differentiation in Machine Learning: A
  Survey}.
\newblock  \bibinfo{volume}{18}, \bibinfo{number}{1} (\bibinfo{date}{Jan.}
  \bibinfo{year}{2017}), \bibinfo{pages}{5595–5637}.
\newblock
\showISSN{1532-4435}


\bibitem[\protect\citeauthoryear{Chadwick}{Chadwick}{1999}]%
        {chadwick}
\bibfield{author}{\bibinfo{person}{Peter Chadwick}.}
  \bibinfo{year}{1999}\natexlab{}.
\newblock \bibinfo{booktitle}{\emph{Continuum mechanics : concise theory and
  problems}}.
\newblock \bibinfo{publisher}{Dover Publications}, \bibinfo{address}{Mineola,
  N.Y}.
\newblock
\showISBNx{978-0486401805}


\bibitem[\protect\citeauthoryear{Ekeland}{Ekeland}{1999}]%
        {convex}
\bibfield{author}{\bibinfo{person}{I Ekeland}.}
  \bibinfo{year}{1999}\natexlab{}.
\newblock \bibinfo{booktitle}{\emph{Convex analysis and variational problems}}.
\newblock \bibinfo{publisher}{Society for Industrial and Applied Mathematics},
  \bibinfo{address}{Philadelphia}.
\newblock
\showISBNx{9780898714500}


\bibitem[\protect\citeauthoryear{Kutz}{Kutz}{2013}]%
        {kutz}
\bibfield{author}{\bibinfo{person}{J.~Nathan Kutz}.}
  \bibinfo{year}{2013}\natexlab{}.
\newblock \bibinfo{booktitle}{\emph{Data-driven modeling \& scientific
  computation : methods for complex systems \& big data}}.
\newblock \bibinfo{publisher}{Oxford University Press},
  \bibinfo{address}{Oxford}.
\newblock
\showISBNx{978-0199660346}


\bibitem[\protect\citeauthoryear{Lecun}{Lecun}{1987}]%
        {lecun1987phd}
\bibfield{author}{\bibinfo{person}{Yann Lecun}.}
  \bibinfo{year}{1987}\natexlab{}.
\newblock \showarticletitle{PhD thesis: Modeles connexionnistes de
  l'apprentissage (connectionist learning models)}.
\newblock  (\bibinfo{year}{1987}).
\newblock


\bibitem[\protect\citeauthoryear{L'Hospital}{L'Hospital}{2015}]%
        {lhospital}
\bibfield{author}{\bibinfo{person}{L'Hospital}.}
  \bibinfo{year}{2015}\natexlab{}.
\newblock \bibinfo{booktitle}{\emph{L'Hospital's analyse des infiniments petits
  : an annotated translation with source material by Johann Bernoulli}}.
\newblock \bibinfo{publisher}{Birkhauser}, \bibinfo{address}{Cham}.
\newblock
\showISBNx{978-3319171142}


\bibitem[\protect\citeauthoryear{Papoulis}{Papoulis}{1962}]%
        {dsp}
\bibfield{author}{\bibinfo{person}{Athanasios Papoulis}.}
  \bibinfo{year}{1962}\natexlab{}.
\newblock \bibinfo{booktitle}{\emph{The Fourier integral and its
  applications}}.
\newblock \bibinfo{publisher}{McGraw-Hill}, \bibinfo{address}{New York}.
\newblock
\showISBNx{978-0070484474}


\bibitem[\protect\citeauthoryear{Paszke, Gross, Massa, Lerer, Bradbury, Chanan,
  Killeen, Lin, Gimelshein, Antiga, Desmaison, Kopf, Yang, DeVito, Raison,
  Tejani, Chilamkurthy, Steiner, Fang, Bai, and Chintala}{Paszke
  et~al\mbox{.}}{2019}]%
        {pytorch}
\bibfield{author}{\bibinfo{person}{Adam Paszke}, \bibinfo{person}{Sam Gross},
  \bibinfo{person}{Francisco Massa}, \bibinfo{person}{Adam Lerer},
  \bibinfo{person}{James Bradbury}, \bibinfo{person}{Gregory Chanan},
  \bibinfo{person}{Trevor Killeen}, \bibinfo{person}{Zeming Lin},
  \bibinfo{person}{Natalia Gimelshein}, \bibinfo{person}{Luca Antiga},
  \bibinfo{person}{Alban Desmaison}, \bibinfo{person}{Andreas Kopf},
  \bibinfo{person}{Edward Yang}, \bibinfo{person}{Zachary DeVito},
  \bibinfo{person}{Martin Raison}, \bibinfo{person}{Alykhan Tejani},
  \bibinfo{person}{Sasank Chilamkurthy}, \bibinfo{person}{Benoit Steiner},
  \bibinfo{person}{Lu Fang}, \bibinfo{person}{Junjie Bai}, {and}
  \bibinfo{person}{Soumith Chintala}.} \bibinfo{year}{2019}\natexlab{}.
\newblock \showarticletitle{PyTorch: An Imperative Style, High-Performance Deep
  Learning Library}. In \bibinfo{booktitle}{\emph{Advances in Neural
  Information Processing Systems}},
  \bibfield{editor}{\bibinfo{person}{H.~Wallach},
  \bibinfo{person}{H.~Larochelle}, \bibinfo{person}{A.~Beygelzimer},
  \bibinfo{person}{F.~d\textquotesingle Alch\'{e}-Buc},
  \bibinfo{person}{E.~Fox}, {and} \bibinfo{person}{R.~Garnett}} (Eds.),
  Vol.~\bibinfo{volume}{32}. \bibinfo{publisher}{Curran Associates, Inc.},
  \bibinfo{pages}{8026--8037}.
\newblock
\urldef\tempurl%
\url{https://proceedings.neurips.cc/paper/2019/file/bdbca288fee7f92f2bfa9f7012727740-Paper.pdf}
\showURL{%
\tempurl}


\bibitem[\protect\citeauthoryear{Qian}{Qian}{1999}]%
        {momentum}
\bibfield{author}{\bibinfo{person}{Ning Qian}.}
  \bibinfo{year}{1999}\natexlab{}.
\newblock \showarticletitle{On the momentum term in gradient descent learning
  algorithms}.
\newblock \bibinfo{journal}{\emph{Neural networks}} \bibinfo{volume}{12},
  \bibinfo{number}{1} (\bibinfo{year}{1999}), \bibinfo{pages}{145--151}.
\newblock


\bibitem[\protect\citeauthoryear{Rall}{Rall}{1981}]%
        {autodiff1981}
\bibfield{author}{\bibinfo{person}{Louis Rall}.}
  \bibinfo{year}{1981}\natexlab{}.
\newblock \bibinfo{booktitle}{\emph{Automatic differentiation : techniques and
  applications}}.
\newblock \bibinfo{publisher}{Springer-Verlag}, \bibinfo{address}{Berlin New
  York}.
\newblock
\showISBNx{978-3-540-10861-0}


\bibitem[\protect\citeauthoryear{Su, Vargas, and Sakurai}{Su
  et~al\mbox{.}}{2019}]%
        {hack}
\bibfield{author}{\bibinfo{person}{Jiawei Su},
  \bibinfo{person}{Danilo~Vasconcellos Vargas}, {and} \bibinfo{person}{Kouichi
  Sakurai}.} \bibinfo{year}{2019}\natexlab{}.
\newblock \showarticletitle{One pixel attack for fooling deep neural networks}.
\newblock \bibinfo{journal}{\emph{IEEE Transactions on Evolutionary
  Computation}} \bibinfo{volume}{23}, \bibinfo{number}{5}
  (\bibinfo{year}{2019}), \bibinfo{pages}{828--841}.
\newblock


\bibitem[\protect\citeauthoryear{Sutskever, Martens, Dahl, and
  Hinton}{Sutskever et~al\mbox{.}}{2013}]%
        {init}
\bibfield{author}{\bibinfo{person}{Ilya Sutskever}, \bibinfo{person}{James
  Martens}, \bibinfo{person}{George Dahl}, {and} \bibinfo{person}{Geoffrey
  Hinton}.} \bibinfo{year}{2013}\natexlab{}.
\newblock \showarticletitle{On the importance of initialization and momentum in
  deep learning}. In \bibinfo{booktitle}{\emph{International conference on
  machine learning}}. \bibinfo{pages}{1139--1147}.
\newblock


\bibitem[\protect\citeauthoryear{Taylor}{Taylor}{1983}]%
        {taylor}
\bibfield{author}{\bibinfo{person}{Angus Taylor}.}
  \bibinfo{year}{1983}\natexlab{}.
\newblock \bibinfo{booktitle}{\emph{Advanced calculus}}.
\newblock \bibinfo{publisher}{Wiley}, \bibinfo{address}{New York}.
\newblock
\showISBNx{978-0471025665}


\end{thebibliography}

\end{document}